\documentclass[twocolumn,twocolumn]{IEEEtran}
\usepackage[T1]{fontenc}
\usepackage{color}
\usepackage{float}
\usepackage{mathrsfs}
\usepackage{amsmath}
\usepackage{amssymb}
\usepackage{graphicx}
\usepackage[unicode=true,
 bookmarks=true,bookmarksnumbered=true,bookmarksopen=true,bookmarksopenlevel=1,
 breaklinks=false,pdfborder={0 0 0},pdfborderstyle={},backref=false,colorlinks=false]
 {hyperref}
\hypersetup{pdftitle={Your Title},
 pdfauthor={Your Name},
 pdfpagelayout=OneColumn, pdfnewwindow=true, pdfstartview=XYZ, plainpages=false}

\makeatletter

\floatstyle{ruled}
\newfloat{algorithm}{tbp}{loa}
\providecommand{\algorithmname}{Algorithm}
\floatname{algorithm}{\protect\algorithmname}

\usepackage[caption=false,font=footnotesize]{subfig}
\usepackage{algorithm}
\usepackage{algorithmic}

\usepackage{multirow} 
\usepackage{amsmath} 
\usepackage{xcolor}

\allowdisplaybreaks[4]

\ifCLASSOPTIONcompsoc
\usepackage[caption=false,font=normalsize,labelfont=sf,textfont=sf]{subfig}
\else
\usepackage[caption=false,font=footnotesize]{subfig}
\fi

\usepackage{cite}
\usepackage{bm}
\usepackage{algorithmic}
\usepackage{algorithm}
\usepackage{graphicx}
\IEEEoverridecommandlockouts
\usepackage[justification=centering]{caption}
\usepackage{lettrine}
\usepackage{geometry}
\@ifundefined{showcaptionsetup}{}{%
 \PassOptionsToPackage{caption=false}{subfig}}
\usepackage{subfig}
\makeatother
\begin{document}

\title{Impact of UAVs Equipped with ADS-B on the Civil
Aviation Monitoring System\\}

\author{\IEEEauthorblockN{Yiyang Liao$^{\dagger}$, Lei Zhang$^{\dagger}$, Ziye Jia$^{\dagger}$, Chao Dong$^{\dagger}$$^{\ast}$, Yifan Zhang$^{\dagger}$, Qihui Wu$^{\dagger}$, Huiling Hu$^{\nparallel}$ and Bin Wang$^{\nparallel}$\\
 }\IEEEauthorblockA{$^{\dagger}$The Key Laboratory of Dynamic Cognitive System of Electromagnetic Spectrum Space, \\
 Ministry of Industry and Information Technology, Nanjing University of Aeronautics and Astronautics \\
 $^{\nparallel}$Middle-south Regional Air Traffic Management Bureau of CAAC\\
$^{\ast}$Corresponding author, email: dch@nuaa.edu.cn}}

\maketitle
\pagestyle{empty}
\thispagestyle{empty}
\begin{abstract}
In recent years, there is an increasing demand for unmanned aerial vehicles (UAVs) to complete multiple applications. However, as unmanned equipments, UAVs lead to some security risks to general civil aviations. In order to strengthen the flight management of UAVs and guarantee the safety, UAVs can be equipped with automatic dependent surveillance-broadcast (ADS-B) devices. In addition, as an automatic system, ADS-B can periodically broadcast flight information to the nearby aircrafts or the ground stations, and the technology is already used in civil aviation systems. However, due to the limited frequency of ADS-B technique, UAVs equipped with ADS-B devices result in the loss of packets to both UAVs and civil aviation. Further, the operation of civil aviation are seriously interfered. Hence, this paper firstly examines the packets loss of civil planes at different distance, then analyzes the impact of UAVs equipped with ADS-B on the packets updating of civil planes. The result indicates that the 1090MHz band blocking is affected by the density of UAVs. Besides, the frequency capacity is affected by the requirement of updating interval of civil planes. The position updating probability within 3s is 92.3$\%$ if there are 200 planes within 50km and 20 UAVs within 5km. The position updating probability within 3s is 86.9$\%$ if there are 200 planes within 50km and 40 UAVs within 5km.
\end{abstract}

\vspace{-0.4cm}
\section{Introduction}
\lettrine[lines=2]{W}{{ith}} the development of low-altitude economics, the unmanned aerial vehicles (UAVs) play increasingly important roles in this area\cite{b1}. Moreover, UAVs are considered as a significant technique in the future wireless access network\cite{b2}\cite{b3}. UAVs can be utilized to many applications, such as surveillance system, aviation photography, information retransmission, etc. Besides, UAVs can be applied to many detailed scenarios such as fire rescue, ecological observation, terrain exploration, etc\cite{b4}. However, how to ensure the safe operation of UAVs becomes an important issue due to the lack of on-board operation. One possible solution is to equip UAVs with automatic dependent surveillance-broadcast (ADS-B) facilities, which is maturely applied in civil aviations\cite{b5}.
\par ADS-B in international civil aviation organization (ICAO) is used to assist traditional primary and secondary radar surveillance\cite{b6}. An aircraft equipped with ADS-B can automatically transmit its position as well as other information to the surrounding aircrafts, vehicles and ground stations (GS)\cite{b7}. These information not only helpful to flying safety but to other aspects like communication network as well\cite{b8}. The ADS-B technique allows aircrafts to identify each other's positions, and automatically maintain safe distance. In addition, ADS-B helps the air traffic controller (ATC) to monitor and control the aircrafts on the route or approaching the airport terminal. In short, ADS-B has the following advantages:
\begin{itemize}
\item ADS-B helps build new air route in remote aeras lacking ground radar stations, such as polar regions, mountains and oceans.
\item ADS-B helps improve the navigation accuracy, reduce the flight safety separation distance and increase the available airspace. 
\item ADS-B helps the planes to climb and descend without the contact with ATC, to avoid the unavailable airspace and flexibly change air route.
\end{itemize}
\par UAVs equipped with ADS-B system contribute to the flying safety, and simplify the UAV control by the air traffic management (ATM). However, this approach has a limitation that the ADS-B system in UAVs and civil aviation share the same frequency band (1090MHz). Due to the limited frequency resources, if many UAVs use this frequency band, the data packets from UAVs and civil planes will have conflicts, resulting in the packets loss and the longer delay of civil aviation related information. It seriously interferes the monitoring of civil aviation by the ATM. Due to the fault tolerance of air route updating interval, some UAVs can access within the same frequency. However, the frequency resource utilization decreases if the number of accessed UAVs is too small. Thus, how to balance the number of UAVs and civil planes becomes an essential problem. 
\par There exist some related works, for example, \cite{b9} points out that due to the low threshold of software defined radio and the openness of 1090MHz, the packets loss in this frequency band becomes increasingly serious. The loss rate of packets 50km away from the GS is as high as 62$\%$. \cite{b10} points out that 1500 aircrafts can be support to send the ADS-B signal only if the channel can be perfectly allocated. However, apart from ADS-B signal (S-mode extended signal), there also exist abundant S-mode and A/C-mode signal in the channel. Hence, the qua-
\begin{figure}[t]
    \centerline{\includegraphics[width=1\linewidth]{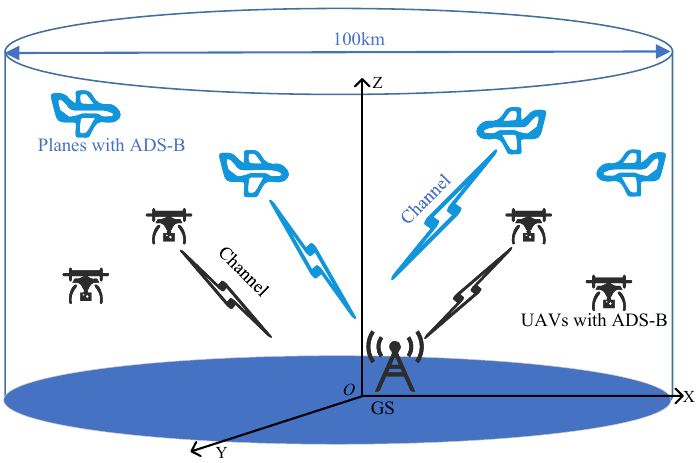}}
    \caption{Considered airspace model.}
    \label{f1}
\end{figure}
\begin{table}[t]
    \caption{\\S-M{\small ODE} E{\small XTENDED} P{\small ACKET}}
    \begin{center}
    \begin{tabular}{|c|c|c|c|c|c|}
    \hline
    Bytes&5bits&3bits&24bits&56bits&24bits \\
    \hline
    Field Name&DF&CA/CF&AA&ME&PI  \\
    \hline
    \end{tabular}
    \label{tab1}
    \end{center}
 \end{table}
 
\noindent ntity of 1500 is only a theoretical value\cite{b11}. In \cite{b12}, the authors use the downlink format (DF) 17 code to represent civil  planes and the DF 18 code to represent UAVs. It points out that the influence of frequency band blocking is mainly affected by the density and the signal transmission power of UAVs. Based on ALOHA protocol, \cite{b13} discusses the relationship between the load and the throughput, and the relationship between the number of planes and the successful receiving probability.\cite{b14} points out that the channel  capacity is mainly affected by the link bit error rate, and changes as the requirement of packets updating interval varies. However, these works don't take the civil planes and UAVs into account at the same time. Also, these works don't consider the 3D model as well as the channel error.
\par As shown in Fig. \ref{f1}, this paper studies the influence of UAVs equipped with ADS-B on 1090MHz and analyzes the infulence of certain number of UAVs accommodated under a given packets updating interval of civil planes. Firstly, this work examines the packets collision loss of civil planes at different distance without UAVs. Then, we consider the error loss into the channel propagation. Finally, we analyze the packets loss of civil planes with different number of UAVs. The data indicates that the 1090MHz band blocking is affected by the density of the UAVs. Besides, the frequency capacity is affected by the requirement of updating interval of civil planes.
\par The rest of this paper is organized as follows. The ADS-B related packets are introduced in Section \uppercase\expandafter{\romannumeral2}. Channel model is introduced in Section \uppercase\expandafter{\romannumeral3}. Section \uppercase\expandafter{\romannumeral4} provides the simulation results and corresponding analyses. Section \uppercase\expandafter{\romannumeral5} draws the conclusions.

\begin{table}[t]
    \caption{\\P{\small ACKET} I{\small NTERVAL}}
    \begin{center}
    \begin{tabular}{|c|c|c|}
    \hline
    ADS-B Packet&Generating interval (s)&Shaking interval (s) \\
    \hline
    POS&0.5&$0.4\sim0.6$\\
    \hline
    VEL&0.5&$0.4\sim0.6$ \\
    \hline
    ID&5&$4.8\sim5.2$\\
    \hline
    AOS&2.5&$2.4\sim2.6$ \\
    \hline
    TSS&1.25&$1.2\sim1.3$ \\
    \hline
    \end{tabular}
    \label{tab2}
    \end{center}
    \end{table}
    
\begin{table}[t]
    \caption{\\A{\small IR} {\small TO} G{\small ROUND} D{\small ATA} L{\small INK}}
    \begin{center}
    \begin{tabular}{|c|c|c|}
    \hline
    \multicolumn{3}{|c|}{Air-ground 95$\%$ probability of position updating interval} \\
    \hline
    Range&3 miles around terminal&5 miles en route \\
    \hline
    Updating interval (s)&3&6 \\
    \hline
    \end{tabular}
    \label{tab3}
    \end{center}
    \end{table}

\section{ADS-B Related Packets}
This Section demonstrates the structure of classic ADS-B packet and the functions of different fields. Then, this Section elaborates the detailed information about airborne position packet. Finally, this Section introduces the shaking interval of all ADS-B affiliated packets.
\subsection{Structure of Classic ADS-B Packet}
ADS-B is the S-mode extended data packet, and its main body length is 112 bits, as shown in Table \ref{tab1} \cite{b15}. The DF field is 5 bits long and is used to distinguish different downlink formats. DF=17 indicates that the S-mode transponder sends ADS-B packets, while DF=18 indicates that the nonS-mode transponder sends ADS-B packets. The code ability/code format (CA/CF) field has different meanings with different DF values. If DF=17, the field is CA, indicating the transponder capability. If DF=18, the field is CF, indicating the encoding format. The aircraft address (AA) field indicates the address of the transmitting device. The message (ME) field contains the ADS-B service data. The parity and identity (PI) field is a downlink field, indicating parity and identity. However, each ADS-B packet has 8 bits control string, so a single ADS-B packet has a total of 120 bits and lasts 120$\mu$s. ADS-B packets are sent periodically by the aircrafts, or the corresponding packets can be sent after being triggered by specific events.
\par Different ME fields respectively contain the aircraft airborne position (POS), aircraft identification and category (ID), airborne velocity (VEL), aircraft operational status (AOS), aircraft target status and other information (TSS). Different ADS-B packets and their corresponding generating interval are shown in Table \ref{tab2}. The average generating interval of POS packet is 0.5s, and the interval of VEL packet is 0.5s. Furthermore, the interval of ID packet is 5s, and the interval of AOS packet is 2.5s. The interval of TSS packet is 1.25s. In order to mitigate conflicts, the generating interval is not a fixed interval, but a shaking interval.

\subsection{Airborne Position Packet}
The ADS-B system broadcasts the POS packets to the outside world with a period of 0.5s, which is the most important packet of all ADS-B afiliated packets. After receiving several POS packets, the GS generates the track of the corresponding aircraft on its radar. When packets are continuously lost, the GS clears the flight path of the aircraft, and the aircraft is deemed to leave its controlled airspace. Due to the impact of packets collision or channel fading, POS packets are lost inevitably. So the position updating interval is introduced to measure the updating quality, which means the GS doesn't need to receive all the POS packets, but to receive a POS packet in a limited time. As shown in Table \ref{tab3}, as for the air-ground data link, the ADS-B receiver at the airport terminal requires a 95$\%$ probability of no more than 3s, for the updating interval of the POS packet of the aircrafts within 3 miles. The ADS-B receiver on the air route requires a 95$\%$ probability of no more than 6s, for the updating interval of the POS packet of the aircrafts within 5 miles\cite{b16}.
\subsection{The shaking interval}
In this paper, we introduce the shaking interval to replace the average generating interval aiming to mitigate packets conflict. As shown in Table \ref{tab2}, the shaking interval of POS packet is [0.4s, 0.6s]. The shaking interval of VEL packet is [0.4s, 0.6s]. The shaking interval of ID packet is [4.8s, 5.2s]. The shaking interval of AOS packet is [2.4s, 2.6s]. The shaking interval of TSS packet is [1.2s, 1.3s]. The S-mode packet (SMAG) payload is 56-bits, and with the 8 bits control string, the actual total length is 64 bits. In \cite{b17}, the authors point out that the ratio of the number of S-mode and ADS-B packets is 1:1, so we assume that each aircraft transmits 5 S-mode packets per second.

\section{Channel Model}
As shown in Fig. \ref{f1}, planes are randomly distributed in the airspace with a radius of 50km. Due to the small size and limited endurance distance, UAVs are randomly distributed in the airspace with a radius of 5km. The ADS-B receiver in GS is located in the two-dimensional geometric center within the airspace range.
\par All planes and UAVs access the channel of 1090MHz by random access, i.e., the ALOHA protocol is adopted as the media access control (MAC) protocol.  
\par All aircrafts are within the one-hop range, i.e., packets do not need to be relayed. If a packet has no error code during the channel transmission and does not overlap with the packets sent by other aircrafts, it can reach the GS and be correctly received.
\par In processing radio signals, the power or amplitude of a signal is usually expressed in decibels (dB). The following formulas use dB to express the signal relationship.

The ADS-B transmission power of the planes are fixed at \emph{P} (dBm) and the ADS-B transmission power of the UAVs are fixed at \emph{p} (dBm). GS is set as a ADS-B receiving station with a receiving sensitivity of \emph{A} (dBm). A random distance parameter $\emph{d}\in (0\rm{km}\sim50\rm{km})$  is generated for each aircraft, and it represents the straight distance from the aircraft to GS. When the considered time is small, it is assumed that each aircraft is quasi-static, i.e., the distance from each aircraft to GS is invariable. All software-defined radio transmissions are considered as line-of-sight. \emph{f} (MHz) represents the working frequency. Then, the propagation \emph{Loss} (dBm) is calculated by the following propagation formula:

\begin{equation}
Loss=32.44+20{\rm lg}\ d+20{\rm lg}\ f.
\end{equation}

ADS-B packets adopt 8 PSK  channel encoding form for 112 bits\cite{b18} \cite{b19}. The input signal power of the demodulator \emph{S} (dBm) is: 
\begin{equation}
S=P-Loss.
\end{equation}

And $S$ must satisfies: $S\geq A$. The gaussian white noise power is \emph{N=$n_0\times$B}. \emph{$n_0$} is the power density of gaussian white noise, and \emph{B} is the bandwidth. Since the ADS-B system operates at 1090MHz, this frequency band belongs to the ultra-high frequency. In the civil aviation system, 962MHz$\sim$1213MHz is used for distance measuring equipment system, and the channel interval is 1MHz. However, 1090MHz is remained for the ADS-B system. Hence, the bandwidth \emph{B}=1MHz. The input signal to noise ratio (SNR) of the demodulator is \emph{r}:
\begin{equation}
    \begin{split}
    r=\frac{S}{N}.
    \end{split}
\end{equation}

\noindent The bit error rate $P_e$ formula of \emph{M}-scale PSK is:
\begin{equation}
\begin{split}
P_e=1-\frac{1}{2\pi } \int_{-\frac{\pi }{M} }^{\frac{\pi }{M} } e^{-r}[1+\sqrt{4\pi r}\cos \theta e^{r(\cos \theta) ^{2}}\\\frac{1}{\sqrt{2\pi } }\int_{-\infty}^{\sqrt{2r\cos \theta } } e^{-\frac{x^{2}}{2} } \,dx   ] \,d\theta.
\end{split}
\end{equation}

\noindent When \emph{M} is 8, the bit error rate $P_e$ formula is approximated as:

\begin{equation}
P_e\approx erfc(\sqrt{r\sin \frac{\pi }{M} } ).
\end{equation}

 In reality, there are three situations for packets loss: packets collision, packets error, and error packet conflicts with intact packet. Packets collision means two packets arrive at the GS simultaneously, which makes the packets discarded. Packets error means some error codes occur during the channel propagation, which also makes the packets discarded. What's more the packet may continue to reach the GS even it has bit errors and conflicts with intact packet, resulting in packets collision. Therefore, it is not possible to simply generate a random factor at the aircraft to judge whether the packet has bit errors and discard it directly. Besides, a random factor cannot be generated at the GS to determine whether the packet has bit errors and discard it directly. Moreover, to guarantee the fairness to all packets, random factor $i\in (0\sim 1)$ is set in the aircraft. And if $i\textless(1-P_e)$, we record it as a good packet in the array. If $i\geq (1-P_e)$, the packet is recorded as a bad packet in the array. Since all packets propagating in the free space, leading to the collisions with each other and generating packets loss, the GS discards bad packets in statistics collection.

\begin{figure}[t]
    \centerline{\includegraphics[width=1\linewidth]{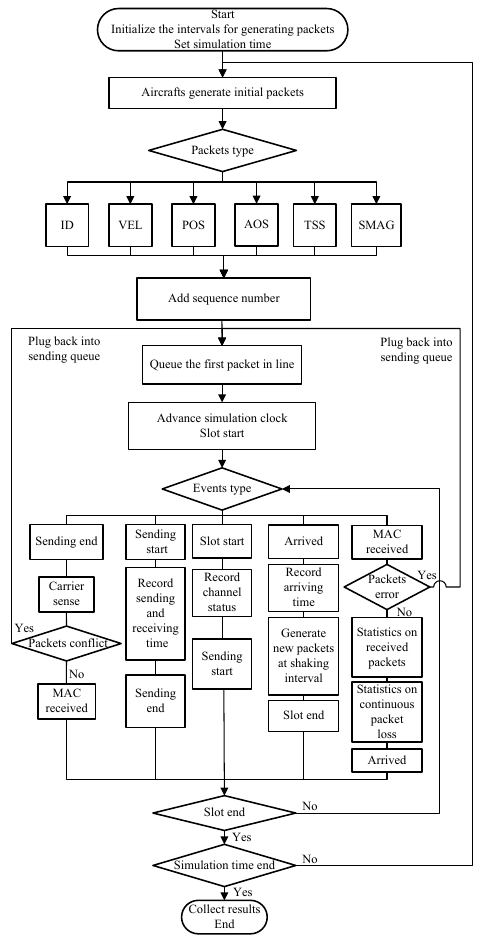}}
    \caption{Flowchart.}
    \label{f2}
\end{figure}

\section{Simulation Results and Analyses}

In this Section, we use CodeBlocks to conduct simulations, aiming to examine the capcity of 1090MHz. The simulations are based on queueing system with discrete events. The events only appear at countable time points, so the system state can only change at countable time points. Without continuity, event processing only needs to be carried out when events change. The system simulation flow chart is shown in the Fig. \ref{f2}. The total simulation time is set as 500 seconds. The operating frequency is 1090MHz.  
\begin{figure}[t]
    \centerline{\includegraphics[width=1.1\linewidth]{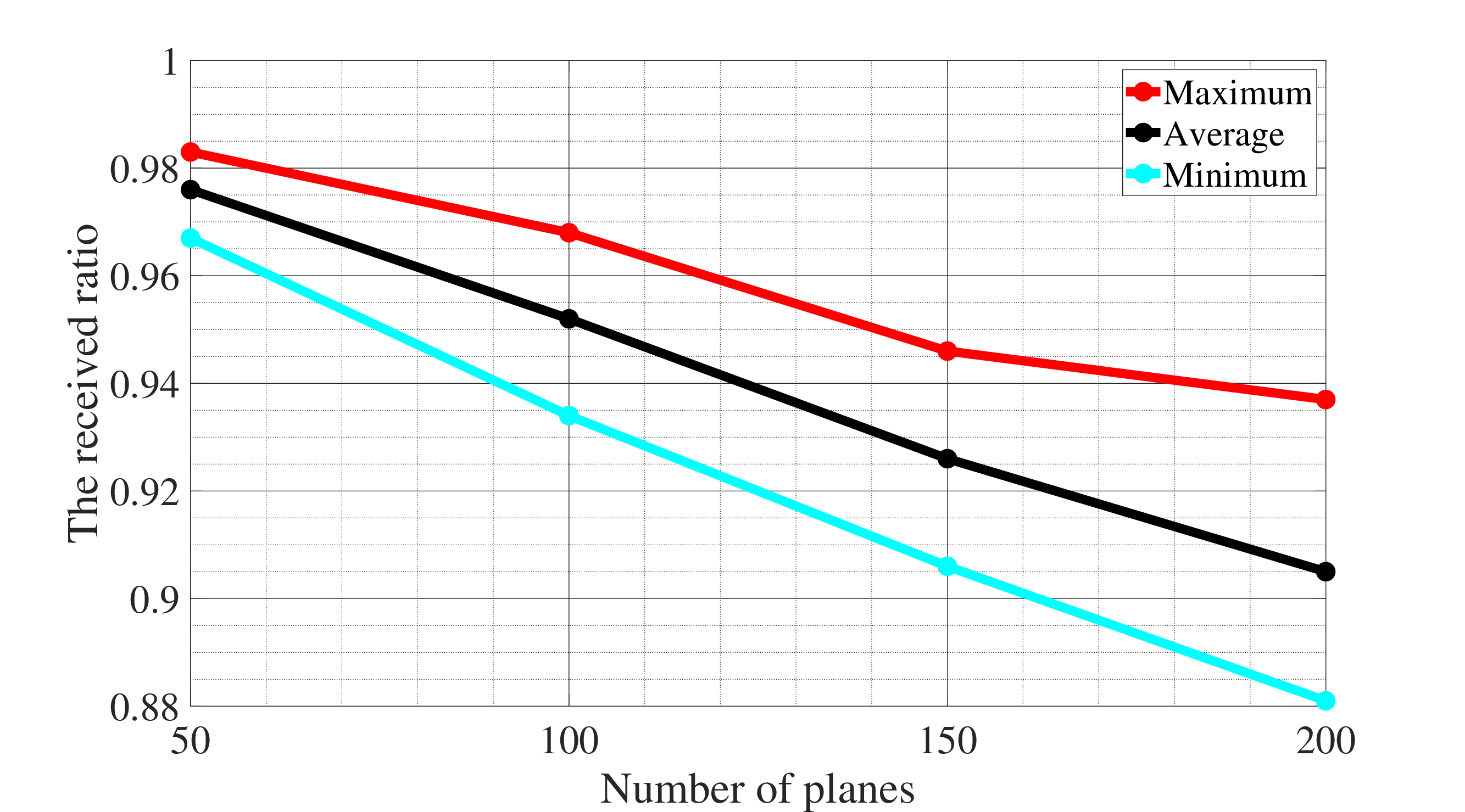}}
    \caption{The received ratio on different planes.}
    \label{f3}
\end{figure}

\noindent The received ratio used in the simulations is defined as:
\begin{equation}
{\rm The\  received\ ratio =\frac{Number\ of\ packets\ received\ by\ GS}{Number\ of\ packets\ sent\ by\ aircrafts} }.
\end{equation}

\par In Fig. \ref{f3}, 50/100/150/200 planes are set respectively, and the ADS-B only sends POS packet and ID packet. There only exists collision packets loss. The number of planes and the received ratio are shown in Fig. 3. As from the Fig. 3, different numbers of planes correspond to different received ratio. As the number of planes increases, the collision of packets intensifies, and the received ratio decreases.

\par
\begin{figure}[t]
    \centerline{\includegraphics[width=1.1\linewidth]{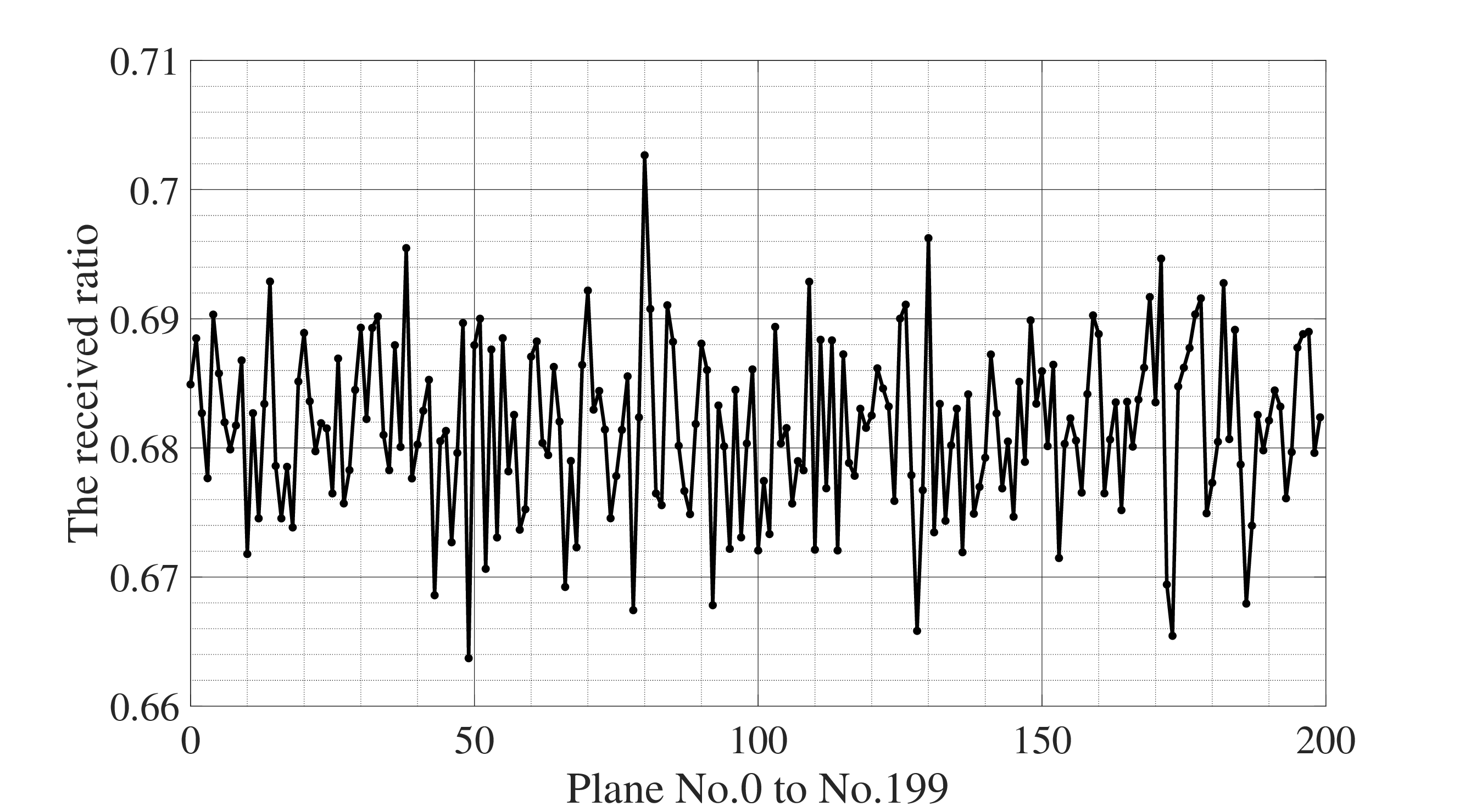}}
    \caption{The received ratio on different planes.}
    \label{f4}
\end{figure}

 As shown in Fig. \ref{f4}, the number of planes is fixed as 200, and the ADS-B sends POS packet, ID packet, VEL packet, TSS packet, AOS packet and S-mode packet. There exists collision packets loss. As shown in Fig. 4, after the introduction of 6 kinds of packets, the received ratio decreases significantly, and the average received ratio is 68.15$\%$. The loss of the POS packet of GS is shown in Table \ref{tab4}. The preceding packets loss cases do not include each other. GS successfully receives 632 POS packets and lost 372 POS packets. The average period of generating POS packet is 0.5s. If the requirement of updating interval is 3s, i.e., more than 6 consecutive packets loss are regarded as the position is not updated. The position updating probability within 3s is 97.8$\%$. 

\par
\begin{figure}[t]
    \centerline{\includegraphics[width=1.1\linewidth]{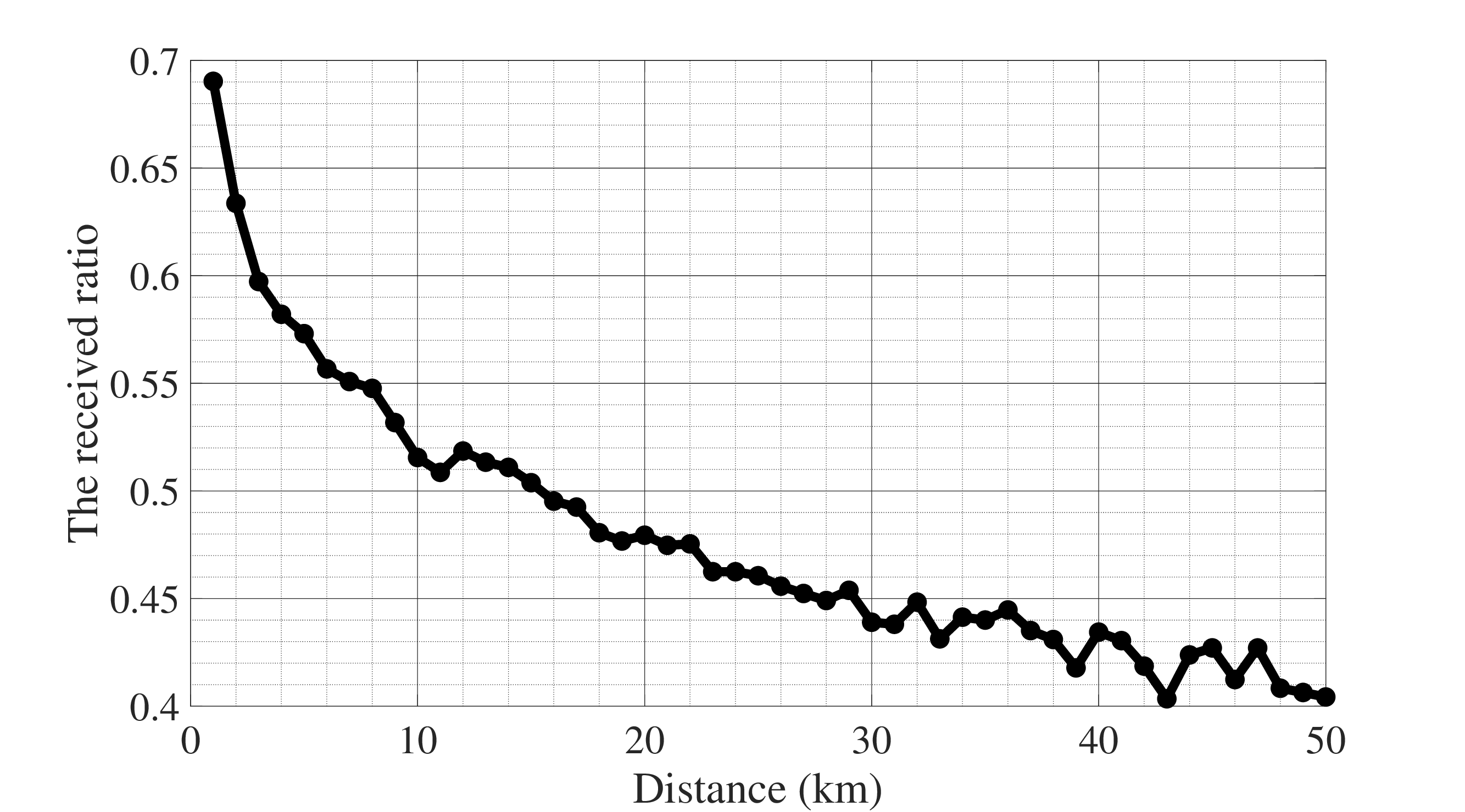}}
    \caption{The received ratio on planes at different distance.}
    \label{f5}
\end{figure}

 As demonstrated in Fig. \ref{f5}, each plane is added with the distance factor, and the packets propagation is added with the channel model. After adding the distance factor, apart from the loss of packets caused by collision, error codes caused by propagation loss also make packets discarded. The number of planes is fixed as 200, and the ADS-B sends 6 kinds of packets. The relationship between the received ratio and distance is shown in Fig. 5. As the distance increases, the propagation loss increases, the packet error increases, and the received ratio decreases. The average received ratio is 48.66$\%$. The loss of the POS packet of GS is shown in Table \ref{tab4}. In total, 464 POS packets are received and 533 POS packets are lost. The average period of generating POS packet is 0.5s. If the requirement of updating interval is 3s, i.e., more than 6 consecutive packets loss are regarded as the position is not updated. The position updating probability within 3s is 93.2$\%$.

\par
\begin{figure}[t]
    \centerline{\includegraphics[width=1.1\linewidth]{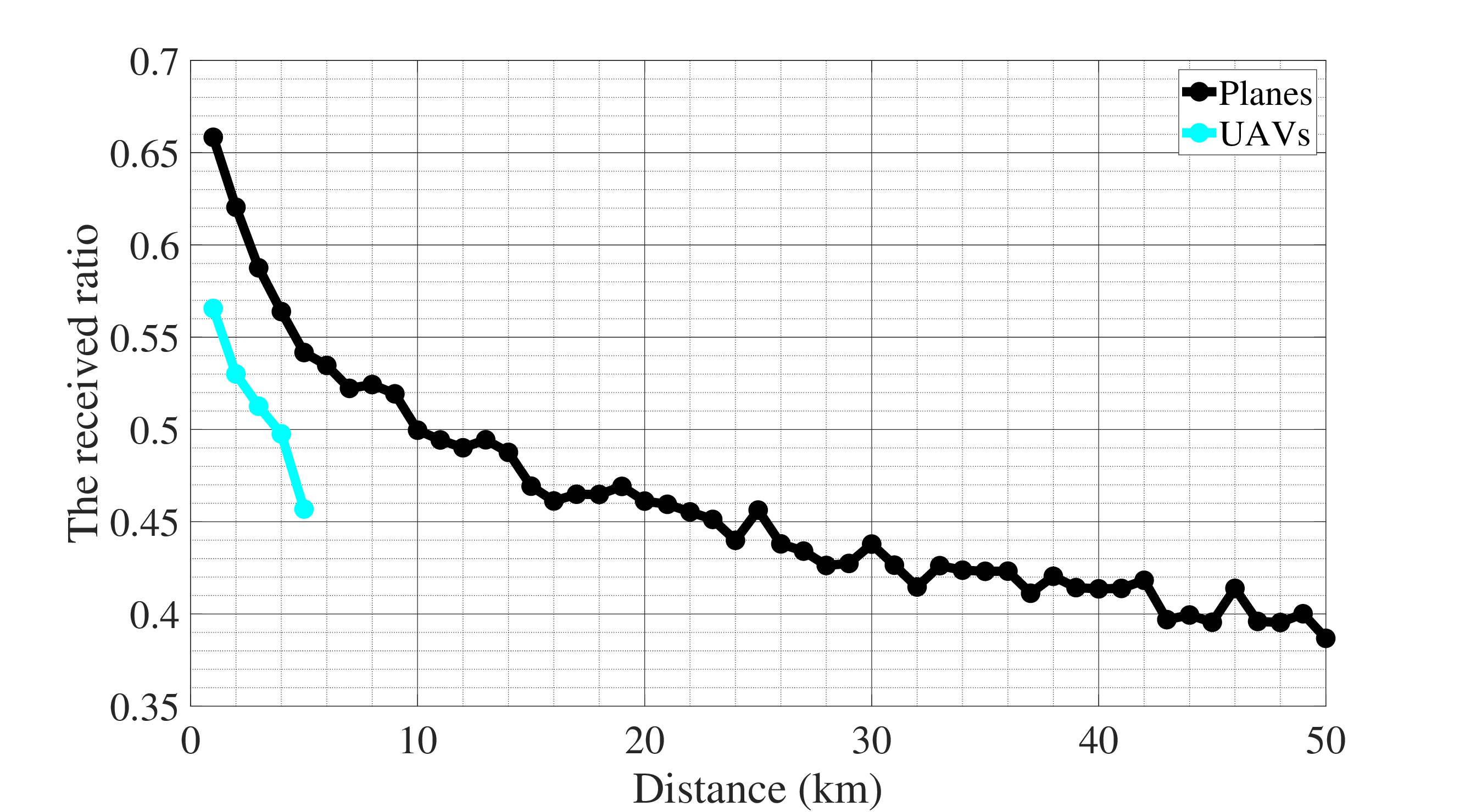}}
    \caption{The received ratio on planes and UAVs at different distance.}
    \label{f6}
\end{figure}

 A total of 200 planes and 20 UAVs are set in Fig. \ref{f6}. The planes are randomly distributed in the 0$\sim$50km airspace range with \emph{P}=44dBm (25W), and the UAVs are randomly distributed in the 0$\sim$5km airspace range with \emph{p}=30dBm (1W). The receiving sensitivity \emph{A} of GS is set as -93dBm. All aircrafts send 6 kinds of packets and there exist collision loss and error loss. The relationship between the received ratio and distance of planes and UAVs is shown in Fig. \ref{f6}. Due to the limited frequency band of 1090MHz, the introduction of UAVs aggravates the packets collision loss of planes. At the same time, due to the limited load of UAVs, the transmission power of ADS-B on UAVs is also smaller than that of planes. Therefore, although the distance between GS and UAVs is shorter, UAVs loss rate is larger than that of civil planes. The average received ratio of GS is 46.55$\%$. The loss of POS packet of GS is shown in Table \ref{tab4}. A total of 579 POS packets are received, and 422 POS packets are lost. If the requirement of updating interval is 3s, i.e., more than 6 consecutive packets loss are regarded as the position is not updated. The position updating probability within 3s is 92.3$\%$.

\par
\begin{figure}[t]
    \centerline{\includegraphics[width=1.1\linewidth]{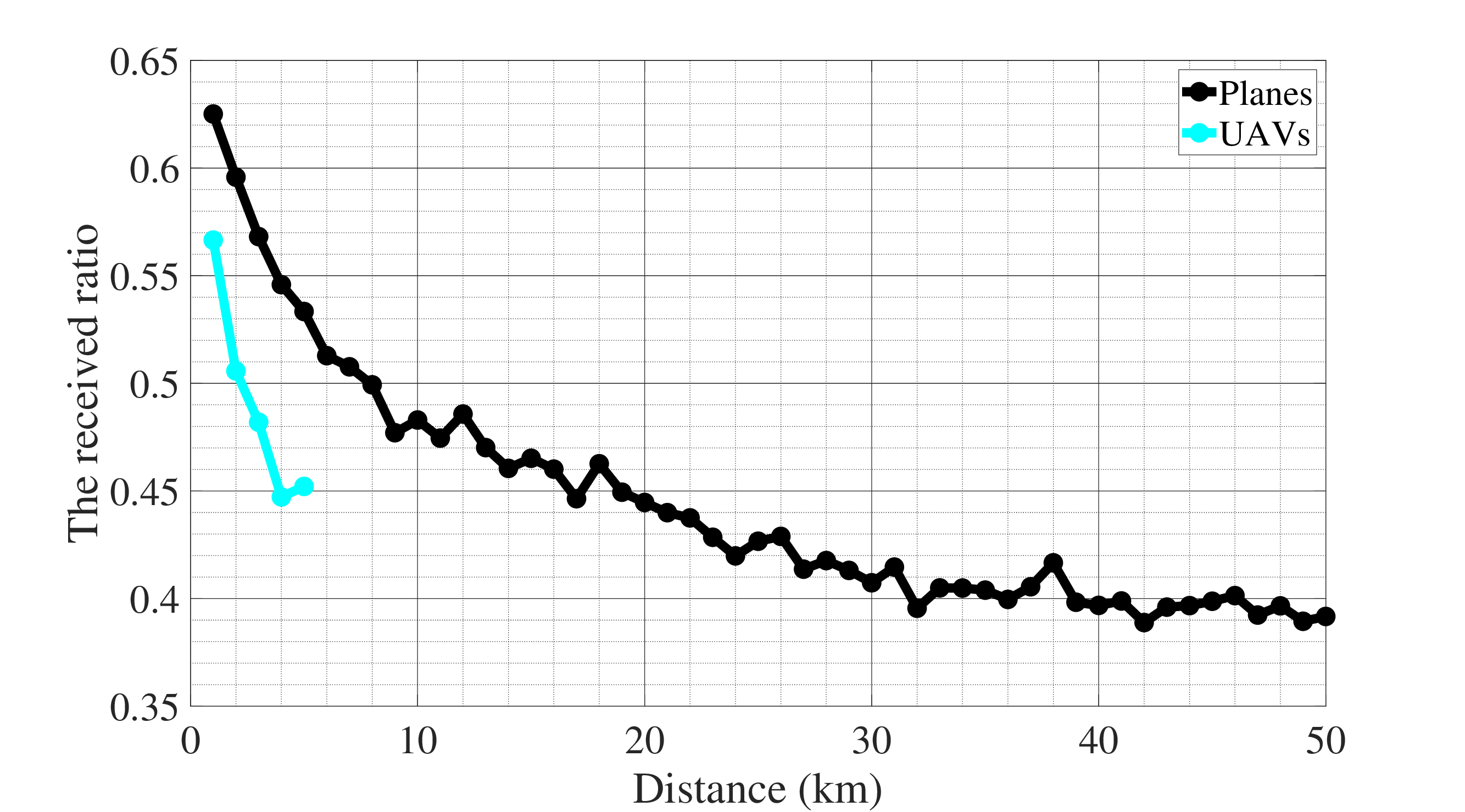}}
    \caption{The received ratio on planes and UAVs at different distance.}
    \label{f7}
\end{figure}

 In Fig. \ref{f7}, a total of 200 planes and 40 UAVs are considered. 

\noindent The planes are randomly distributed in the 0$\sim$50km airspace range with \emph{P}=44dBm (25W), and the UAVs are randomly distributed in the 0$\sim$5km airspace range with \emph{p}=30dBm (1W). The receiving sensitivity \emph{A} of GS is set as -93dBm. All aircrafts send 6 kinds of packets and there exist collision loss and error loss. The relationship between the received ratio and distance of planes and UAVs is shown in Fig. 7. The average received ratio of the GS is 45.25$\%$. The loss of the POS packet of GS is shown in Table \ref{tab4}. A total of 364 POS packets are received and 629 POS packets are lost. If the requirement of updating interval is 3s, i.e., more than 6 consecutive packets loss are regarded as the position is not updated. The position updating probability within 3s is 86.9$\%$.

\begin{table*}[t]
    \caption{\\P{\small ACKETS} L{\small OSS} {\small OF} A{\small IRBORNE} P{\small OSITION}}
    \begin{center}
    \begin{tabular}{|c|c|c|c|c|}
    \hline
    N packets are lost consecutively&Times (GS in Fig. 4)&Times (GS in Fig. 5)&Times (GS in Fig. 6)&Times (GS in Fig. 7)  \\
    \hline
    1 packet&144&82&134&95 \\
    \hline
    2 packets&50&49&40&46 \\
    \hline
    3 packets&25&36&24&36 \\
    \hline
    4 packets&5&14&8&22 \\
    \hline
    5 packets&1&11&3&11 \\
    \hline
    6 packets&1&11&2&10 \\
    \hline
    7 packets&2&5&1&6 \\
    \hline
    8 packets&1&3&3&2\\
    \hline
    9 packets&0&1&4&4\\
    \hline
    10 packets&0&0&1&1\\
    \hline
    11 packets&0&0&0&1\\
    \hline
    16 packets&0&0&0&1\\
    \hline
     \end{tabular}
    \label{tab4}
    \end{center}
\end{table*}

\section{Conclusions}
This work analyses the impact of UAVs equipped with ADS-B on the civial planes at the frequency of 1090MHz. A couple of conditions are set up. Firstly, we assume that only civil planes sending 2 kinds of packets access the 1090MHz, as the number of planes increases, the received ratio decreases. The packets may be discarded only due to collision loss. Then, we add 6 kinds of packets, which accords with the actual situation. Furthermore, we introduce the channel model, which means the packets may be discarded due to the error loss. Finally, we add the UAVs into the simulation, examining the received ratio and the position updating probability. The simulation data indicates that the 1090MHz band blocking is affected by the density of the UAVs. Besides, the frequency capacity is affected by the requirement of updating interval of civil planes. As the distance increases, the received ratio decreases. The loss ratio of planes 50km away from GS can  approximately reach 60$\%$. Moreover, when there are 200 planes in the 50km radius airspace and 20 UAVs in the 5km radius airspace, the average received ratio of GS is 46.55$\%$ and the position updating probability is 92.3$\%$. If the number of UAVs changed to 40, the average received ratio of GS is 45.25$\%$ and the position updating probability is 86.9$\%$.
\bibliographystyle{IEEEtran}

\vspace{12pt}

\end{document}